\documentclass[aps,prl,twocolumn,superscriptaddress]{revtex4-1}
\usepackage{graphicx}
\usepackage{amsmath}
\usepackage{amsfonts}
\usepackage{amssymb}
\usepackage{pdfpages}

\begin{document}
\title{Non-trivial role of interlayer cation states in iron-based superconductors}

\author{Daniel Guterding}
\email{guterding@itp.uni-frankfurt.de}
\affiliation{Institut f\"ur Theoretische Physik, Goethe-Universit\"at Frankfurt,
Max-von-Laue-Stra{\ss }e 1, 60438 Frankfurt am Main, Germany}

\author{Harald O. Jeschke}
\affiliation{Institut f\"ur Theoretische Physik, Goethe-Universit\"at Frankfurt,
Max-von-Laue-Stra{\ss }e 1, 60438 Frankfurt am Main, Germany}

\author{I. I. Mazin}
\affiliation{Code 6393, Naval Research Laboratory, Washington, DC 20375, USA}

\author{J. K. Glasbrenner}
\altaffiliation[Present address: ]{Department of Computational and Data Sciences
and Computational Materials Science Center, George Mason University, 4400
University Drive, Fairfax, VA 22030}
\affiliation{National Research Council/Code 6393, Naval Research Laboratory,
Washington, DC 20375, USA}

\author{E. Bascones}
\affiliation{Instituto de Ciencia de Materiales de Madrid, ICMM-CSIC, Cantoblanco, 28049 Madrid, Spain}

\author{Roser Valent\'i}
\affiliation{Institut f\"ur Theoretische Physik, Goethe-Universit\"at Frankfurt,
Max-von-Laue-Stra{\ss }e 1, 60438 Frankfurt am Main, Germany}

\begin{abstract}
Unconventional superconductivity in iron pnictides and chalcogenides has been
suggested to be controlled by the interplay of low-energy antiferromagnetic
spin fluctuations and the particular topology of the Fermi surface in these
materials. Based on this premise, one would also expect the large class of
isostructural and isoelectronic iron germanide compounds to be good
superconductors. As a matter of fact, they, however, superconduct at very low
temperatures or not at all. In this work we establish that superconductivity
in iron germanides is suppressed by strong ferromagnetic tendencies, which
surprisingly do not originate from changes in bond-angles or -distances with
respect to iron pnictides and chalcogenides, but are due to changes in the
electronic structure in a wide range of energies happening upon substitution of atom species (As by Ge
and the corresponding spacer cations). Our results indicate that
superconductivity in iron-based materials may not always be fully understood based
on $d$ or $dp$ model Hamiltonians only.
\end{abstract}

\maketitle

\textit{Introduction.-} After the initial discovery of high-temperature
superconductivity in doped LaFeAsO~\cite{Kamihara2008}, a large variety of
other iron pnictides and chalcogenides have been shown to be
superconductors~\cite{Hosono2015a}, with some reports of the transition
temperature $T_{c}$ as high as 100~K~\cite{Ge2015}. On the other hand,
isoelectronic and isostructural iron germanides are either
non-superconducting~\cite{Avila2004, Ran2011, Kim2015, Liu2012} or possibly
superconduct at very low temperatures~\cite{Zou2014, Chen2016}. The currently
most intensively debated material is YFe$_{2}$Ge$_{2}$, for which
superconductivity below 2~K has been reported~\cite{Chen2016}. Its electronic
structure is very similar to that of CaFe$_{2}$As$_{2}$ in the collapsed
tetragonal phase, but with significant hole-doping~\cite{Chen2016, Singh2014,
Subedi2014}. This led to speculation~\cite{Chen2016} about a connection
between superconductivity in YFe$_{2}$Ge$_{2}$ and the collapsed phase of the
extremely hole-doped pnictide, KFe$_{2}$As$_{2}$~\cite{Ying2015, Nakajima2015,
Guterding2015B}. Furthermore, Wang \textit{et al.}~\cite{Wang2016B} recently
found YFe$_{2}$Ge$_{2}$ to be close to a magnetic instability and X-ray
absorption and photoemission experiments show evidence for strong
spin-fluctuations~\cite{Sirica2015} and moderate correlation
effects~\cite{Xu2016} in this material.

\begin{figure}[b]
\includegraphics[width=\linewidth]{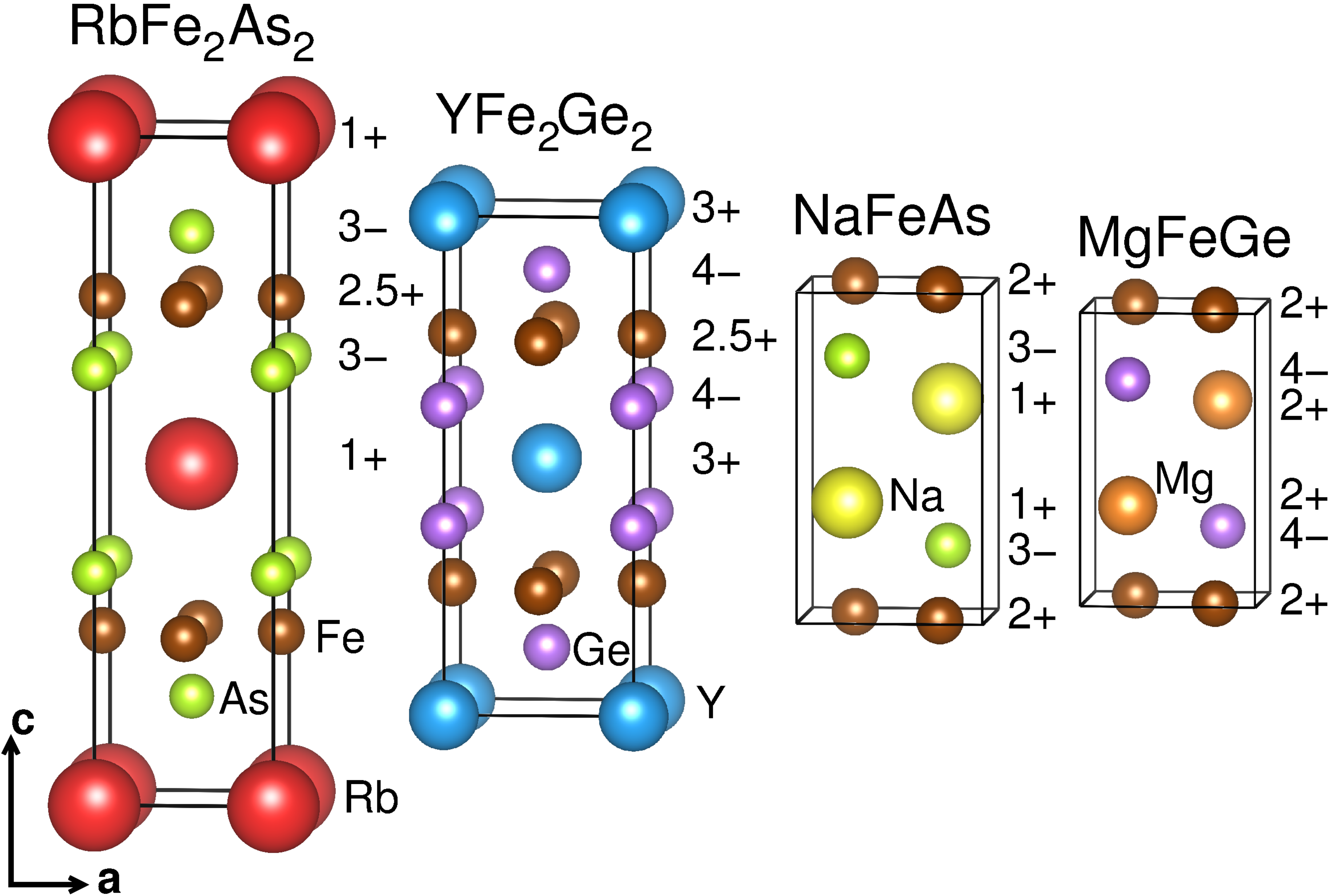}\caption{(Color online) Crystal structures of RbFe$_{2}$As$_{2}$, YFe$_{2}$Ge$_{2}$, NaFeAs and
MgFeGe. The unit cells and interatomic distances are true to scale. The
numbers next to the unit cells indicate the nominal valence of atoms at the
same vertical positions.}
\label{fig:structure}
\end{figure}

It is generally agreed that magnetism plays an important role~in
superconductivity of Fe-based superconductors (FeBS) \cite{Hirschfeld2011,
Chubukov2012, Hosono2015a, Glasbrenner2015, Guterding2016, Davis2013, Hu2016,
Si2016}. It is therefore natural to ask whether the magnetic tendencies in
iron germanides are fundamentally different from those in iron pnictides and
chalcogenides~\cite{Bascones2016} and why that is the case. In a first attempt to understand the
lack of superconductivity in Fe germanides, a few authors investigated the
electronic properties of the isoelectronic and isostructural materials MgFeGe
and LiFeAs~\cite{Jeschke2013,Yin2014,Ding2014}. The former is a paramagnetic
metal, while the latter is a superconductor. An important conclusion was that
the dominant magnetic exchange interactions in MgFeGe are ferromagnetic, while
those in LiFeAs are antiferromagnetic. The microscopic origin of this
different behavior was, however, not further explored.

In this Letter we show that (i) the presence of ferromagnetic tendencies is a general trait in iron
germanides, which is detrimental for superconductivity, and that (ii) the
ferromagnetic tendencies arise from the interaction of the cation spacer
with the FeGe layer. In fact, the hole-doping or collapse of the $c$-axis in
YFe$_{2}$Ge$_{2}$ are not essential for this behavior, but the key is in 
substitution of As by Ge and the corresponding substitution of monovalent or
divalent spacers by divalent or trivalent cations, respectively. This  modifies the
electronic bandstructure in a wide range of energies at and away from the Fermi level
and creates ferromagnetic tendencies which
suppress superconductivity. Hence, one can go from As  to Se/Te, $i.e.,$
right in the periodic table, and find further FeBS, but not to the left
towards Ge. In agreement with recent NMR measurements~\cite{Wiecki2015}, our
study highlights the role of presence or absence of ferromagnetic fluctuations in determining the
value of $T_{c}$ in FeBS.

Our analysis shows that conventional low-energy models of FeBS, which only
incorporate the Fe $d$ and $X$ ($X$=As, Se, Ge, ...)  $p$ states 
 are in some cases not  sufficient to explain  key features of FeBS. Although these models
usually reproduce the Fermi surface very well, they do not reflect the
physical instabilities of the actual materials because they neglect the
interaction with the spacer between the Fe$X$  layers.
Even though  bulk FeSe does not contain spacer layers, our arguments may be
relevant for intercalates~\cite{BurrardLucas2013, Guterding2015A,Sun2015}, alkali-dosed thick films~\cite{Wen2016} and  
FeSe monolayers on SrTiO$_3$~\cite{Ge2015}

\textit{Materials and Methods.-} We compare isoelectronic iron arsenides and
iron germanides from (i) the so-called  hole-doped \textit{122-family} 
where iron is
in a nominal oxidation Fe$^{2.5+}$ with $d^{5.5}$ 
occupation~\cite{Yin2011, Backes2014,
Backes2016} and (ii) the so-called \textit{111-family} with
 Fe$^{2+}$ in a  $d^{6}$ configuration~\cite{Yin2011, Ferber2012, Lee2012}.
 The crystal structures of
RbFe$_{2}$As$_{2}$, YFe$_{2}$Ge$_{2}$, NaFeAs and MgFeGe are shown in
Fig.~\ref{fig:structure}, where we also indicate the nominal valences of the
atoms in each compound. Lattice constants and internal positions in this figure were taken
from experiment~\cite{Venturini1996, Parker2009, Liu2012, Eilers2015}.

The most obvious structural difference between iron
arsenides and iron germanides is shrinking of the $c$-axis
(Fig.~\ref{fig:structure}). From NaFeAs to MgFeGe it is not as pronounced as
from RbFe$_{2}$As$_{2}$ to YFe$_{2}$Ge$_{2}$, where Ge $p_{z}$-$p_{z}$ bonds
may form (in MgFeGe direct Ge-Ge bonding is not possible). Although these
materials are isoelectronic, the germanides have a stronger charge transfer
between the Fe$X$ ($X$= As, Ge) and the spacer layers.

The isoelectronic substitution of As by Ge, Rb by Y, and Na by
Mg was simulated within the virtual crystal approximation (VCA). To
disentangle effects originating from direct atomic substitution from effects
coming from small changes of
bond-distances and -angles in real materials, we performed all calculations
for the \textit{122-family} with the experimental structural parameters of YFe$_{2}$%
Ge$_{2}$~\cite{Venturini1996} and those for the \textit{111-family} with the
experimental structural parameters of MgFeGe~\cite{Liu2012}. The technical details
of our DFT calculations can be found in Ref.~\onlinecite{Supplement}. 

We also analyze the density of states by using the extended Stoner
model~\cite{Andersen1977, Mazin1997}, which is a simple tool for understanding
the origin of itinerant ferromagnetism (see Ref.~\onlinecite{Supplement} for
details). The paramagnetic state is unstable towards ferromagnetism if the
conditions $1/I=\bar{N}(m)$ and $0>d\bar{N}(m)/dm$ are fullfilled at some $m$,
where $\bar{N}(m)$ is the paramagnetic density of states averaged over an
energy window that contains a sufficient number of states to realize an Fe
moment $m,$ and $I$ is the Stoner parameter~\cite{Supplement}. 

\begin{figure}[tb]
\includegraphics[width=\linewidth]{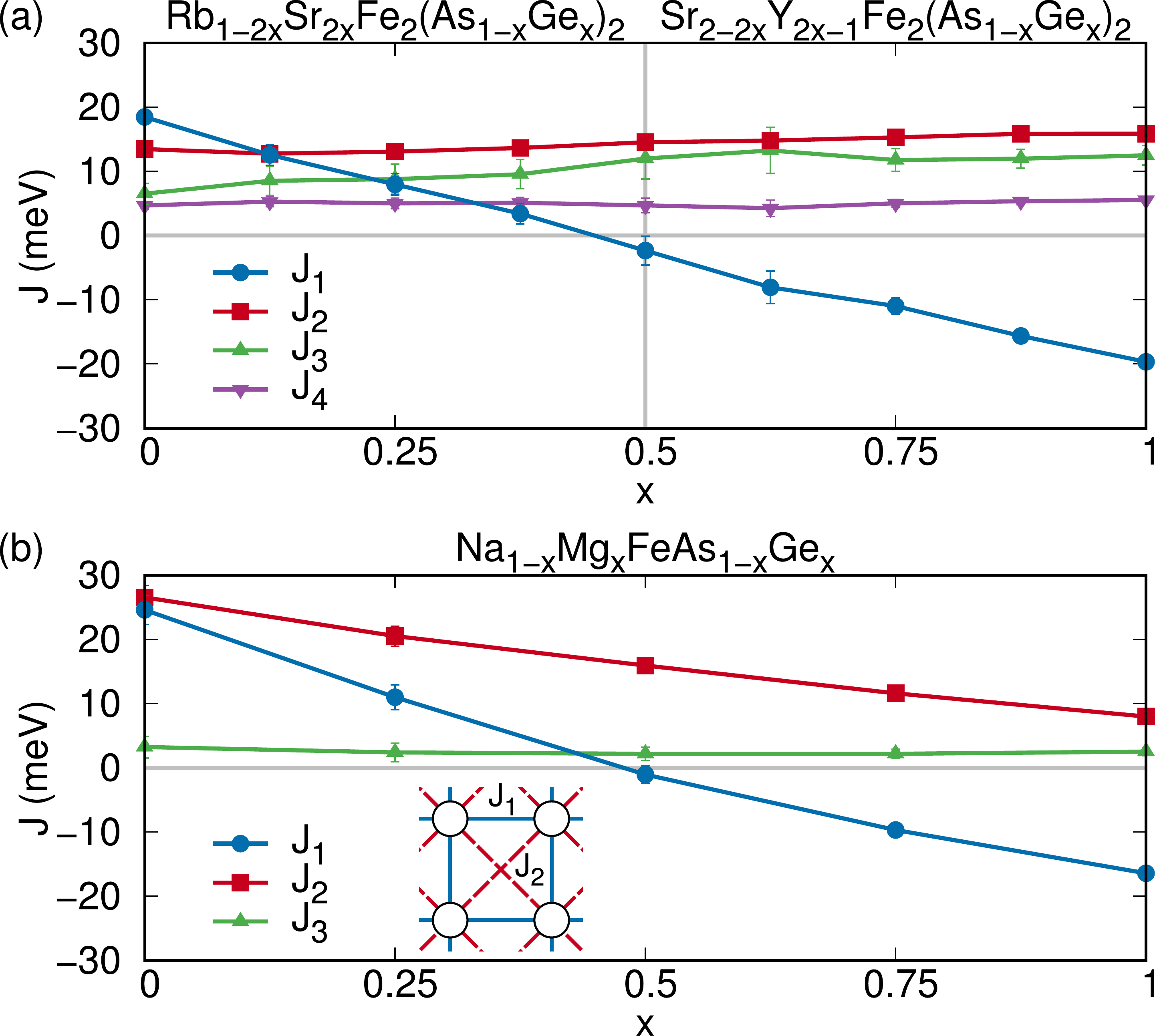}\caption{(Color online) Calculated
Heisenberg exchange parameters for (a) the VCA interpolation between
RbFe$_{2}$As$_{2}$ and YFe$_{2}$Ge$_{2}$ [via SrFe$_{2}$(As$_{0.5}$Ge$_{0.5}%
$)$_{2}$] and (b) the VCA interpolation between NaFeAs and MgFeGe. Lines are
guides to the eye. The error bars represent the statistical errors of the fit.
The inset of (b) shows the structure of the two-dimensional Heisenberg model we use to fit the DFT energies. $J_{1}$ is the nearest-neighbor coupling in the square lattice
of Fe atoms, while $J_{2}$ is the next-nearest neighbor coupling. $J_3$ and $J_4$ are longer-range exchange couplings. Positive
values of J correspond to antiferromagnetic exchange. Note that all
calculations were performed in the crystal structures of YFe$_{2}$Ge$_{2}$ and
MgFeGe respectively.}
\label{fig:heisenberg}
\end{figure}

\textit{Results.-} We first calculated the DFT energies of various spin configurations. By means of the VCA we interpolated between RbFe$_{2}$As$_{2}$ and YFe$_{2}$Ge$_{2}$ [via SrFe$_{2}$(As$_{0.5}$Ge$_{0.5}$)$_{2}$] and between NaFeAs and MgFeGe. Using a two-dimensional Heisenberg model to parametrize the DFT energies (see
Ref.~\onlinecite{Supplement} for more details) we observe that the
nearest-neighbor exchange coupling $J_{1}$ universally changes from
antiferromagnetic to ferromagnetic when going continuously from As to Ge without
changing the electron count, while all other exchange couplings are almost unaffected (Fig.~\ref{fig:heisenberg}).  Only
in the \textit{111-family} the next-nearest-neighbor exchange $J_{2}$ is also
reduced, but it does not change sign. At the germanide end-point the ferromagnetic $J_{1}$ becomes the dominant exchange interaction.

Remarkably, we also obtained a large ferromagnetic $J_1$ for NaFeAs
  after we
expanded the structure used for Fig.~\ref{fig:heisenberg} by 10\% along the
$c$-axis but kept  all distances within the FeAs layer unchanged by the
expansion. These results indicate that NaFeAs can also be turned ferromagnetic by separating
the FeAs layers and by shifting Na further away from the layers. 

From this analysis we conclude that previous suggestions~\cite{Wang2016B} that
iron germanides and  iron pnictides show similar magnetic behavior don't hold.
While both families have a stripe antiferromagnetic ground state
in the DFT calculations,
the nature of excitations is entirely different. This is
reflected in the presence of a nearest neighbor ferromagnetic exchange $J_1$ 
 in  iron germanides
 and antiferromagnetic $J_1$ in the iron pnictides despite the 
very similar crystal structure and electronic structure at the Fermi level.
In particular, the results on the expanded NaFeAs  suggest
that the origin of this different behavior lies dominantly in the relative separation
between the spacer and the Fe$X$ plane.

A further distinctive feature of the germanides
is that the magnetism of Fe in YFe$_{2}$Ge$_{2}$ 
appears to be rather peculiar. There is a
low- and a high-moment solution for Fe, the former more stabilized for shorter
Fe-Ge bond length~\cite{Supplement} (in pnictides, either a high-spin solution
is realized, or magnetism collapses completely).

\begin{figure}[tb]
\includegraphics[width=\linewidth]{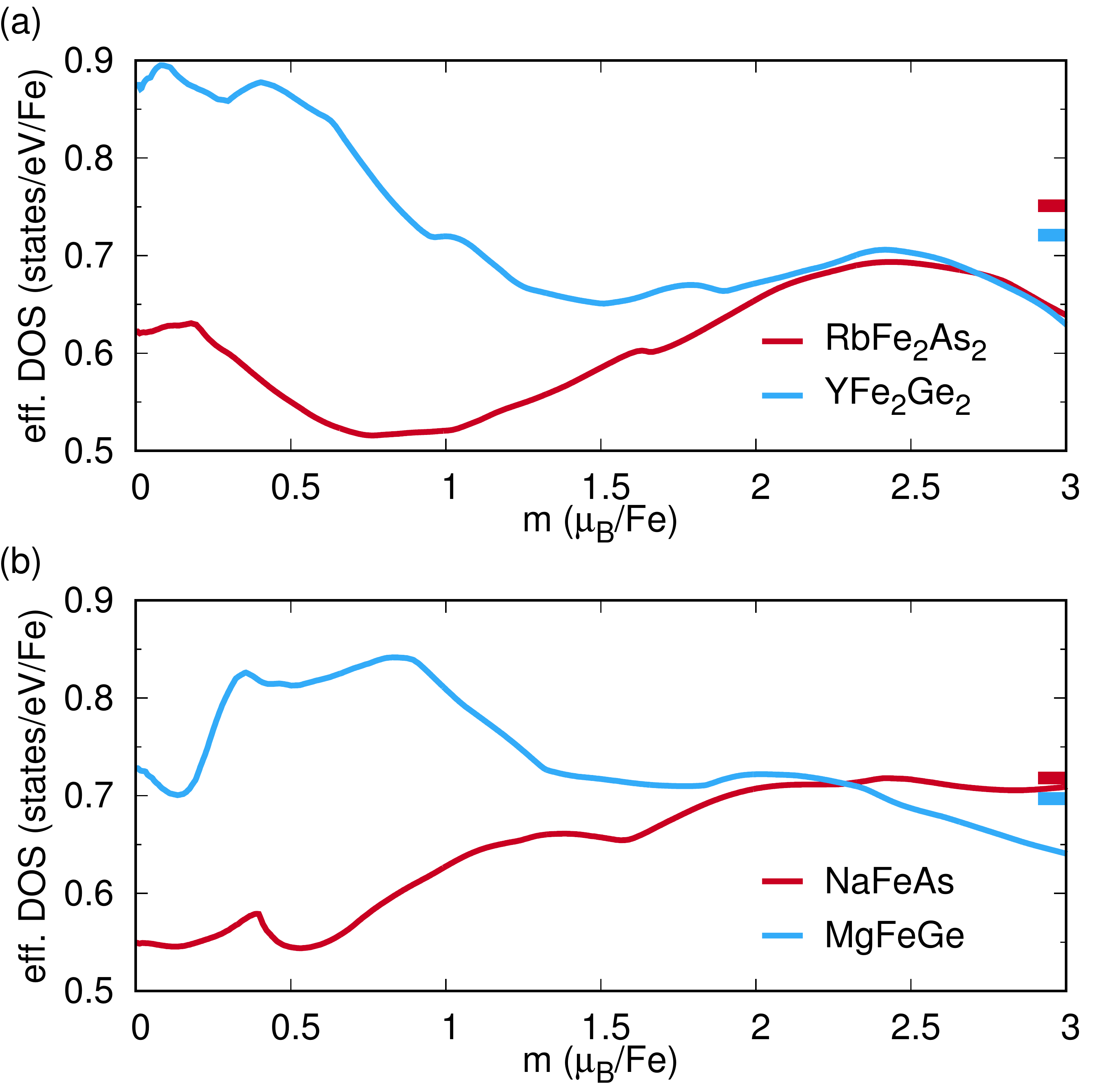}\caption{(Color online) Effective
density of states in the extended Stoner model as a function of magnetic
moment for (a) RbFe$_{2}$As$_{2}$ and YFe$_{2}%
$Ge$_{2}$ and
(b) NaFeAs and MgFeGe. The colored bars on the
right $y$-axis indicate the calculated inverse Stoner parameters $1/I$ for the
respective case. All calculations were performed in the crystal structures of
YFe$_{2}$Ge$_{2}$ and MgFeGe respectively.}
\label{fig:stonereffdos}
\end{figure}

\begin{figure}[tb]
\includegraphics[width=\linewidth]{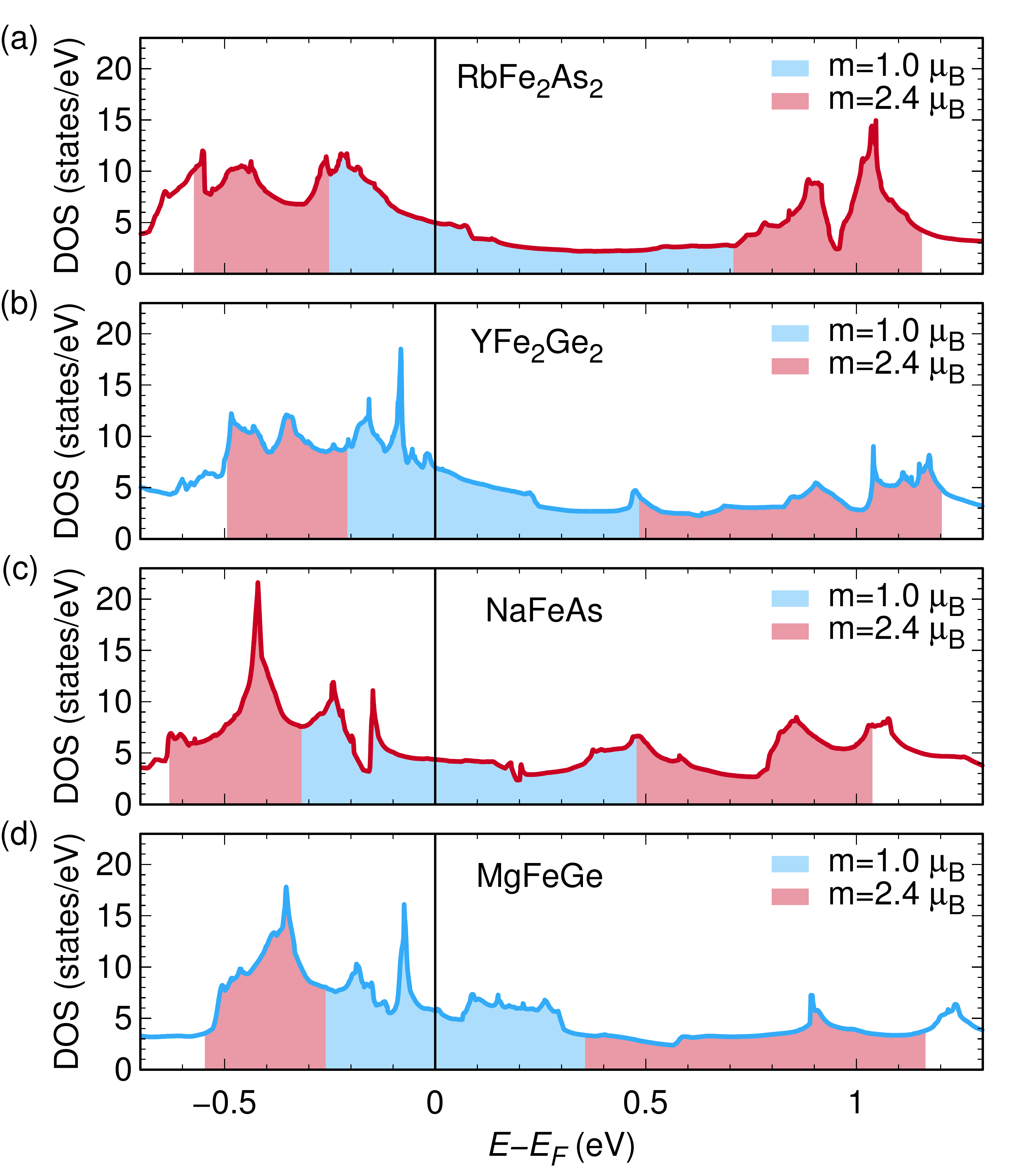}\caption{(Color online) Total density
of states calculated from DFT for (a) RbFe$_{2}$As$_{2}$, (b) YFe$_{2}$%
Fe$_{2}$, (c) NaFeAs and (d) MgFeGe. The shaded areas below the curves
correspond to the energy range needed to realize a moment of $1.0\,\mu_{B}$ or
$2.4\,\mu_{B}$ per iron respectively within the extended Stoner model. All
calculations were performed in the crystal structures of YFe$_{2}$Ge$_{2}$ and
MgFeGe respectively.}
\label{fig:dos}
\end{figure}

To understand in a simple framework the origin of
the magnetic behavior
 presented above we investigate the effective density of states $\bar{N}$ as a
function of the magnetic moment $m$ within the extended
 Stoner model (see Fig.~\ref{fig:stonereffdos}).
We observe that (i) iron germanides have in general a higher DOS at the Fermi
level and (ii) a significant number of states is shifted from higher energies
towards the Fermi level, as compared to pnictides.  This is signalled by the strong
increase of the effective DOS at low moments (see Fig.~\ref{fig:stonereffdos}
where results for YFe$_{2}$Ge$_{2}$ versus RbFe$_{2}$As$_{2}$ and MgFeGe
versus NaFeAs are shown). Interestingly, the changes in the high-moment region
($m\sim2.4~\mu_{B}$) are marginal, while they are considerable in the
low-moment region ($m<1.0~\mu_{B}$). Furthermore, we find that the Stoner
parameter $I$ is almost independent of the material and that $1/I$ lies
between 0.7~eV$^{-1}$ and 0.75~eV$^{-1}$. Therefore, by looking for crossings
of $\bar{N}(m)$ with $1/I$ in Fig.~\ref{fig:stonereffdos}, we establish that
the extended Stoner criterion for ferromagnetism is fulfilled in iron
germanides, but not in pnictides. Moreover, the metastability of different
magnetic moments in YFe$_{2}$Ge$_{2}$ is also evident from this analysis, as
the effective DOS almost fulfills the extended Stoner criterion also for large
moments of about $2.5~\mu_{B}$.

Fig.~\ref{fig:dos} shows the total calculated DOS for RbFe$_{2}$As$_{2}$ vs.
YFe$_{2}$Ge$_{2},$ and NaFeAs vs. MgFeGe, where we colored the energy regions
corresponding to magnetic moments of $m=1.0~\mu_{B}$ (blue) and $m=2.4~\mu
_{B}$ (red) in the extended Stoner model. The energy range corresponding to
$m=1.0~\mu_{B}$ is compressed when going from arsenides to germanides, while
the energy range corresponding to $m=2.4~\mu_{B}$ even increases marginally in
germanides. As the density of states in the window shown is dominated by Fe
states, this implies that the bandwidth of some of the Fe states is
selectively reduced in iron germanides, while the overall bandwidth is about constant~\cite{Supplement}.

\textit{Discussion.-} One of the principal questions in the theory of the
Fe-based superconductors is what should be the minimal chemical model to explain the
essential physics and, above all, superconductivity. It was recognized 
 that the effective Fe-only (\textquotedblleft $d$-only\textquotedblright)
model does not work in some materials, but it has been believed so far that the
electronic properties of iron-based superconductors were exclusively controlled
by the Fe$X$ layers ($X$=As, Se, Ge, ...) as described  by the
so-called \textquotedblleft $pd$-model\textquotedblright.
Thereby the role of all other constituents was reduced to charge
reservoirs.

We have established in this work that iron germanides have a general tendency towards
 ferromagnetism which proves detrimental for superconductivity even
though the Fermi surface is very similar to that of isoelectronic pnictides.
Most importantly, this tendency can be traced down to the flattening of some bands
near the Fermi level and  a modified
electronic bandstructure in a wide range of energies at and away from the Fermi level.
 Neither the collapse of the $c$-axis, nor the
hole-doping of the \textit{122} germanides are essential for the emergence of
ferromagnetism. However, the character and position of the intercalating
species, normally considered irrelevant and not explicitly included in any theory or
model, plays a decisive role.

Our findings have important implications for iron-based superconductivity in general:
(i) The Fermi surface geometry and topology is an important, but not the only
condition for emerging superconductivity. The character of spin fluctuations,
even on the level of the simple ferromagnetic-antiferromagnetic dichotomy, may
be qualitatively different in seemingly similar materials.
(ii) A quantitative theory of $T_{c}$ in iron-based superconductors must
include the interaction between all constituents of the unit cell, including,
in some cases, the interlayer spacers.
(iii) While FeGe layers $per$ $se$ are not necessarily ferromagnetic, the
fact that they have to be spaced with different elements (\textit{e.g., }Mg
vs. Na, or Y vs. Sr) drives them ferromagnetic.
(iv) In a more general way, it does matter what we place next to or on top of
an Fe-ligand layer. This observation may be directly related to an apparent
role that interfacial effects play in high-$T_c$ Fe chalcogenides, such as FeSe
monolayers deposited on specially prepared surfaces or K$_{x}$Fe$_{2-y}$Se$_{2}$
filaments embedded in the magnetic K$_{2}$Fe$_{4}$Se$_{5}$ phase.

\begin{acknowledgments}
\textit{Acknowledgments.-} DG, HOJ and RV thank the German Research Foundation
(Deutsche Forschungsgemeinschaft) for financial support through grant SPP
1458. IIM was supported by ONR through the NRL basic research program and by the
Alexander von Humboldt foundation. JKG acknowledges the support of the NRC program
at NRL. EB acknowledges funding from Ministerio de Econom\'ia y Competitividad v\'ia
Grant No. FIS2014-53219-P and from Fundaci\'on Ram\'on Areces and 
thanks R. Rurali and X. Cartoixa for early calculations.  
The authors thank S. L. Bud'ko and P. C.
Canfield for helpful discussions.
\end{acknowledgments}

\clearpage


\section*{Supplementary material}

# Supplemental Information: Non-trivial role of interlayer cation states in iron-based superconductors

Daniel Guterding,[1, *] Harald O. Jeschke,[1] I. I. Mazin,[2] J. K. Glasbrenner,[3, †] E. Bascones,[4] and Roser Valentí[1]

[1] Institut für Theoretische Physik, Goethe-Universität Frankfurt,
Max-von-Laue-Straße 1, 60438 Frankfurt am Main, Germany
[2] Code 6393, Naval Research Laboratory, Washington, DC 20375, USA
[3] National Research Council/Code 6393, Naval Research Laboratory, Washington, DC 20375, USA
[4] Instituto de Ciencia de Materiales de Madrid, ICMM-CSIC, Cantoblanco, 28049 Madrid, Spain

## I. METASTABILITY OF MAGNETIC STATES WITH DIFFERENT ORDERED MOMENTS IN $YFe_2Ge_2$

We analyzed the energies of different magnetic configurations in $YFe_2Ge_2$ using the FPLO density functional theory code[1] with a GGA exchange-correlation functional[2], following Refs. 3–5. We investigate the GGA+U ordered moment of $YFe_2Ge_2$ as a function of the Fe-Ge bond length, starting with $U = 0$. For the magnetic order we choose ferromagnetic arrangement within the iron plane and antiferromagnetic stacking along the $c$-axis. We find that $YFe_2Ge_2$ shows a peculiar metastability of ordered magnetic states with ordered moments of $m \sim 1\mu_B$ (low moment) and $m \sim 2\mu_B$ (high moment). The ground state for small Fe-Ge bond lengths has an ordered moment of about $m \sim 1\mu_B$, while the $m \sim 2\mu_B$ state can be stabilized for larger Fe-Ge bond lenghts. The results are summarized in Fig. 1. Note that some of the solutions initialized as high moment actually converged to a low moment solution.

Furthermore, we investigated the influence of Coulomb repulsion on the Fe 3$d$ orbitals. We find that even a tiny Coulomb repulsion $U = 0.25$ eV is sufficient to make the high moment compete for the global energy minimum. The

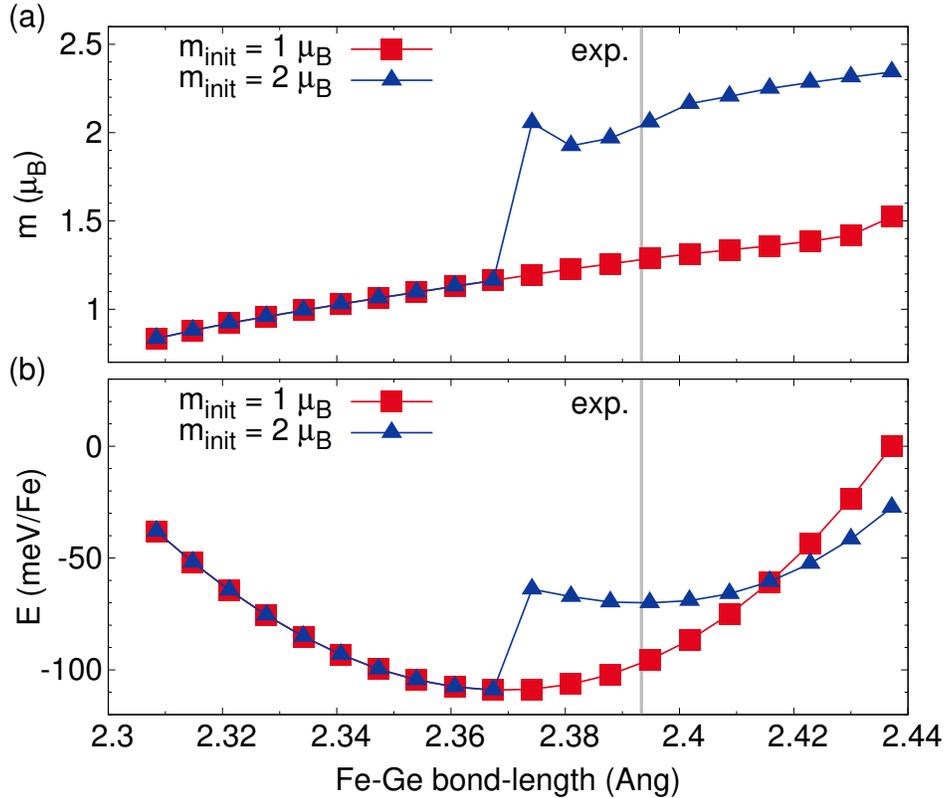

FIG. 1. (a) Magnetic moment versus Fe-Ge bond length and (b) total energy versus Fe-Ge bond length for $YFe_2Ge_2$. All calculations were performed with GGA xc-functional. The red curve was generated from ferromagnetic initial configurations within the Fe plane with a moment of $1\mu_B$ per Fe. The ferromagnetic planes were stacked antiferromagnetically in the $c$-direction. Calculations with an initial magnetic moment of $2\mu_B$ per Fe result in the blue curve. The vertical grey line marks the experimental Fe-Ge bond-length.

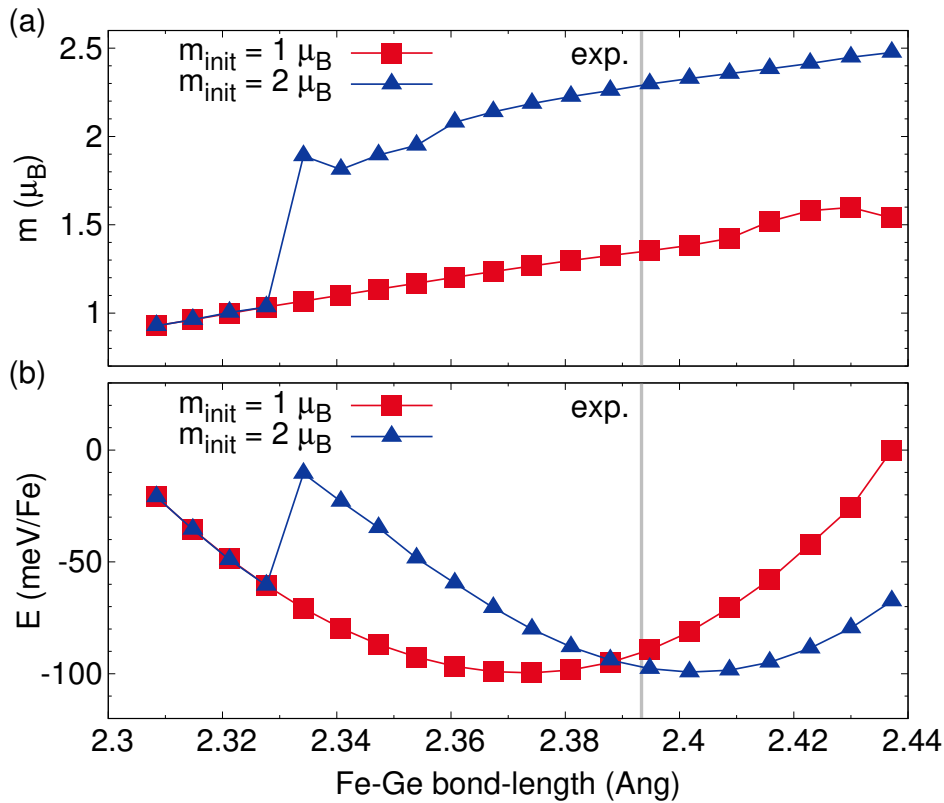

FIG. 2. (a) Magnetic moment versus Fe-Ge bond length and (b) total energy versus Fe-Ge bond length for YFe$_2$Ge$_2$. All calculations were performed with GGA+U xc-functional using $U = 0.25$ eV. The red curve was generated from ferromagnetic initial configurations within the Fe plane with a moment of $1\mu_B$ per Fe. The ferromagnetic planes were stacked antiferromagnetically in the c-direction. Calculations with an initial magnetic moment of $2\mu_B$ per Fe result in the blue curve. The vertical grey line marks the experimental Fe-Ge bond-length.

results for this calculation are summarized in Fig. 2. Note that a larger Coulomb repulsion also shifts the minimum in the energy for the high-moment solutions to larger Fe-Ge bond lengths.

The metastability of low- and high-moment solutions lead us to the investigation of the germanide compounds within the extended Stoner formalism.

## II. METASTABILITY OF MAGNETIC STATES WITH DIFFERENT ORDERED MOMENTS IN MgFeGe

Using the Wien2k code and otherwise comparable calculation setup we also investigated the energy of the low- and high-moment solution in MgFeGe as a function of the Ge height. Depending on the Ge height, either the low- or the high-spin solution is stable, and the other solution manifests itself as a shoulder (inflection point) in the $E(M)$ curve. This behavior is shown in Fig. 3, which shows the values of the magnetic moment $M$ at the energy minimum and the inflection point as a function of the Ge $z$-position.

## III. EXTENDED STONER MODEL

In the extended Stoner calculation we follow Ref. 6. The extended Stoner theory determines whether the paramagnetic state of a material is stable against ferromagnetism.

Within the extended Stoner theory the total magnetization energy can be written as Eq. 1, where $m$ is the magnetization and $\bar{N}(m)$ is the density of states averaged between the Fermi levels of the spin-up and the spin-down channels. In the limit of $m \to 0$ the effective Stoner DOS is simply the total DOS at the Fermi level. For all other moments $m$ the effective Stoner DOS is the average total DOS within an increasingly large energy window around the Fermi

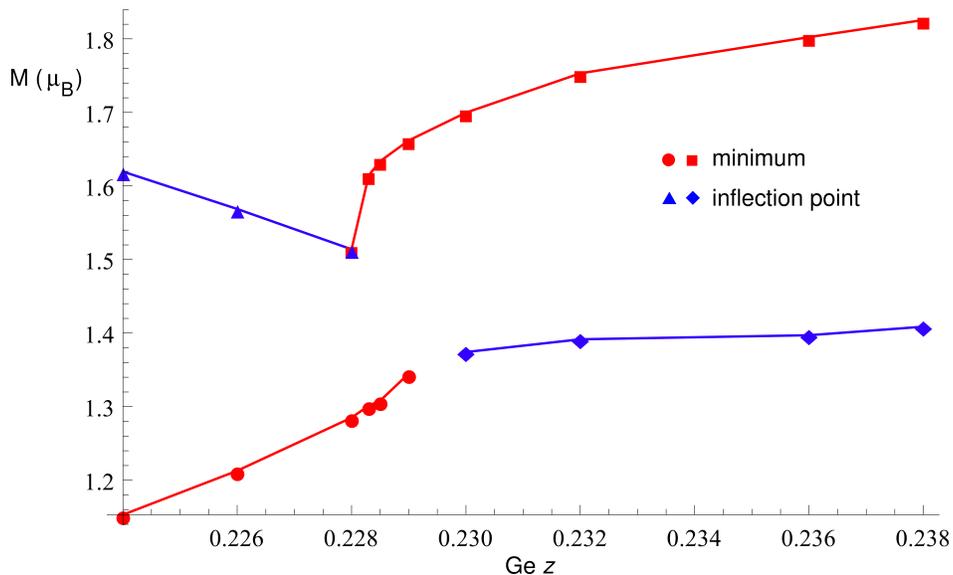

FIG. 3. Magnetic moment $M$ at the minimum and the inflection point of the $E(M)$ curve from fixed-moment calculations in MgFeGe as a function of the Ge $z$-poition (Ge height). The energy minimum changes from the low-spin to the high-spin solution when Ge height is increased beyond a certain value. Lines are guides to the eye.

energy. The density of states is evaluated within the rigid-band approximation starting from the paramagnetic state.

$$E(m) = \frac{1}{2} \int_0^m \frac{m' dm'}{\bar{N}(m')} - \frac{Im^2}{4} \quad (1)$$

The first term is the one-electron energy, because the change in energy for a single electron upon generating a moment $m$ is half the energy difference $\Delta(m)$ between the Fermi levels of the spin-up and spin-down channels. Using $\Delta(m)\bar{N}(m) = m$ one obtains Eq. 2. Integrating this expression one obtains the first term in Eq. 1.

$$\frac{dE}{dm} = \frac{\Delta(m)}{2} = \frac{m}{2\bar{N}(m)} \quad (2)$$

The second term is the so-called Stoner parameter, which parametrizes the ferromagnetic exchange-interactions between electrons in an averaged form.

Now we derive the conditions under which the paramagnetic state is unstable towards ferromagnetism. We start by finding the extrema in the energy for Eq. 1. These are given by Eq. 3.

$$0 \stackrel{!}{=} \frac{dE}{dm} = \bar{N}(m)^{-1} - I \quad (3)$$

Next we find the condition under which this extremum is a minimum in the magnetization energy. It is given by Eq. 4, where we used $I = \bar{N}(m)^{-1}$.

$$0 \stackrel{!}{<} \frac{d^2E}{dm^2} = \frac{1}{2\bar{N}(m)} - m\frac{d\bar{N}(m)/dm}{2\bar{N}(m)^2} - \frac{I}{2} = -m\frac{d\bar{N}(m)/dm}{2\bar{N}(m)^2} \quad (4)$$

As all moments $m$ and effective densities of states are positive, the resulting condition is $0 > d\bar{N}(m)/dm$.

Combining Eqs. 3 and 4 we find that the paramagnetic state is unstable towards ferromagnetism if the conditions $1 = \bar{N}(m)I$ and $0 > d\bar{N}(m)/dm$ are fullfilled for the same value of $m$.

The Stoner parameter $I$ can be obtained from fixed-moment DFT calculations by inserting the DFT energies for $E(m)$ into Eq. 1, using the paramagnetic DFT density of states to calculate $\bar{N}(m)$ and performing a numerical optimization to obtain the value of $I$ on the parabolic term in Eq. 1. The effective Stoner parameter $I$ was calculated from a fit to DFT total energies of ferromagnetic configurations with a moment of up to 3 $\mu_B$ per iron site. All calculations for the Stoner analysis were converged using $20^3$ $k$-point grids.

The results are shown in Fig. 4. The values for the Stoner parameters are 1.331 eV (RbFe$_2$As$_2$ in YFe$_2$Ge$_2$ structure), 1.388 eV (YFe$_2$Ge$_2$), 1.392 eV (NaFeAs in MgFeGe structure) and 1.435 (MgFeGe). Note that the fit is good in the high moment region. The deviation for low moments probably originates from an admixture of non-Fe states, which increases the effective Stoner parameter. This would also explain why the fit is generally worse for germanide compounds, where also a significant amount of interlayer cation states is at the Fermi level. As we use extended Stoner theory only as a qualitative tool, these discrepancies are of minor importance.

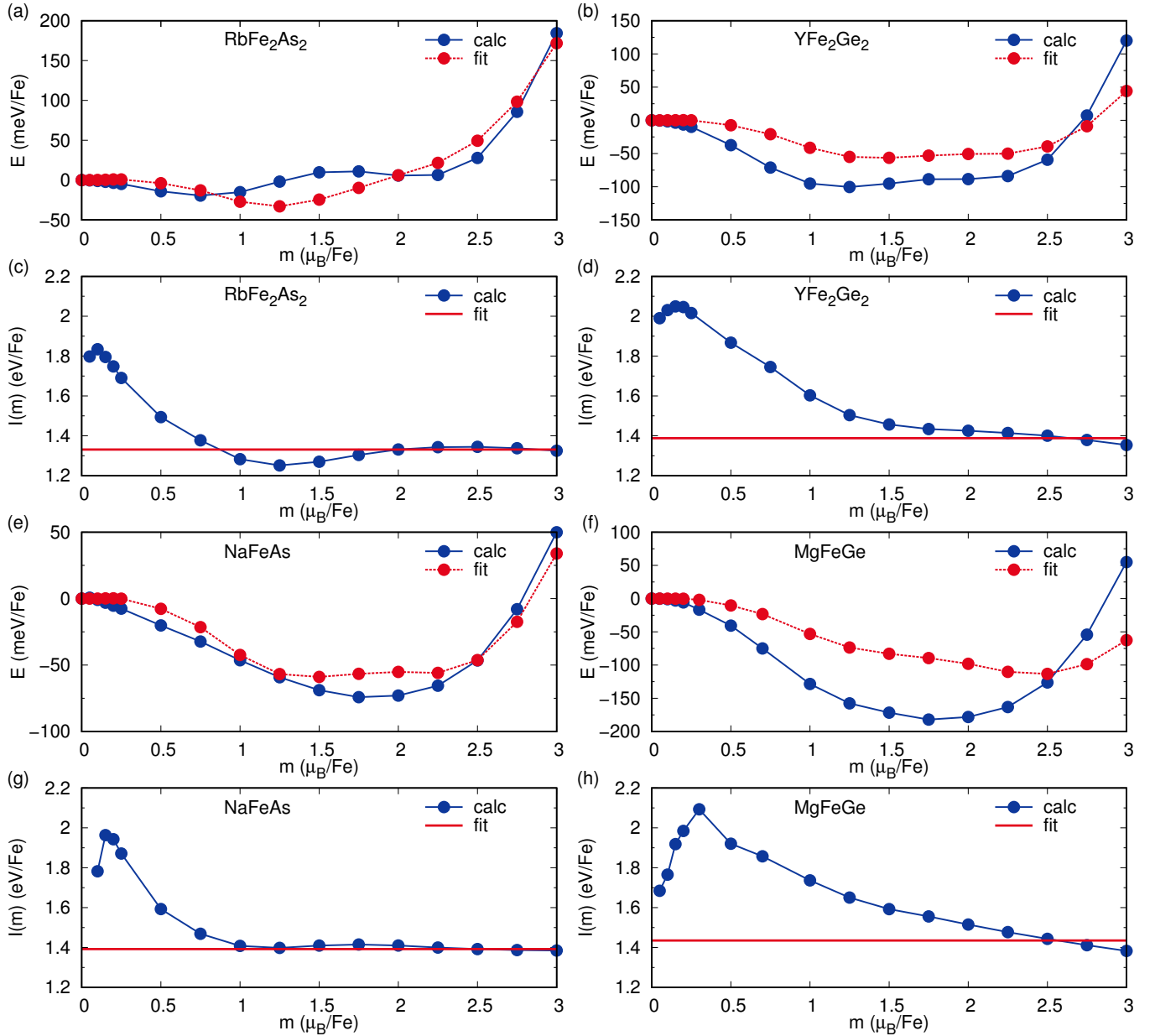

FIG. 4. (a),(b),(e),(f) Fit of Eq. 1 to the fixed-moment energies calculated from DFT. (c),(d),(g),(h) Same data as in (a),(b),(e),(f) after subtracting the one-electron energies (first term in Eq. 1) and dividing by $-m^2/4$.

## IV. DETAILS ON CALCULATION OF HEISENBERG EXCHANGE PARAMETERS

We calculate the magnetic exchange interactions based on a $2 \times 2 \times 1$ supercell containing eight Fe atoms. This leads to 17 inequivalent magnetic configurations for the *122-compounds* and 13 inequivalent configurations for the *111-compounds*. We used $8^3$ $k$-point grids for converging these magnetic calculations. The Heisenberg exchange couplings were extracted by mapping DFT total energies of all inequivalent magnetic configurations to a classical Heisenberg model for each composition.

## V. ELECTRONIC STRUCTURE FOR VCA INTERPOLATION BETWEEN NaFeAs AND MgFeGe

We investigate the changes in electronic structure when continously turning MgFeGe into NaFeAs using the virtual crystal approximation, while keeping all structural parameters constant. The electronic bandstructures with orbital weights are shown in Figs. 5, 6 and 7.

The flattening of bands described in the main text can be seen in Fig. 5, where the Fe $d_{xz}$ states at the $X$-point are lowered almost to the Fermi level in MgFeGe and the corresponding band becomes very flat. In the same figure one can see that the Fe $d_{xy}$ states in turn increase their bandwidth when going from NaFeAs to MgFeGe, for example by comparing the path segment $M$-$\Gamma$. Fig. 7 shows that the cation states are lowered very much in energy, for example at the $M$-point. Furthermore, their weight at the lower end of the energy range shown increases when going from NaFeAs to MgFeGe.

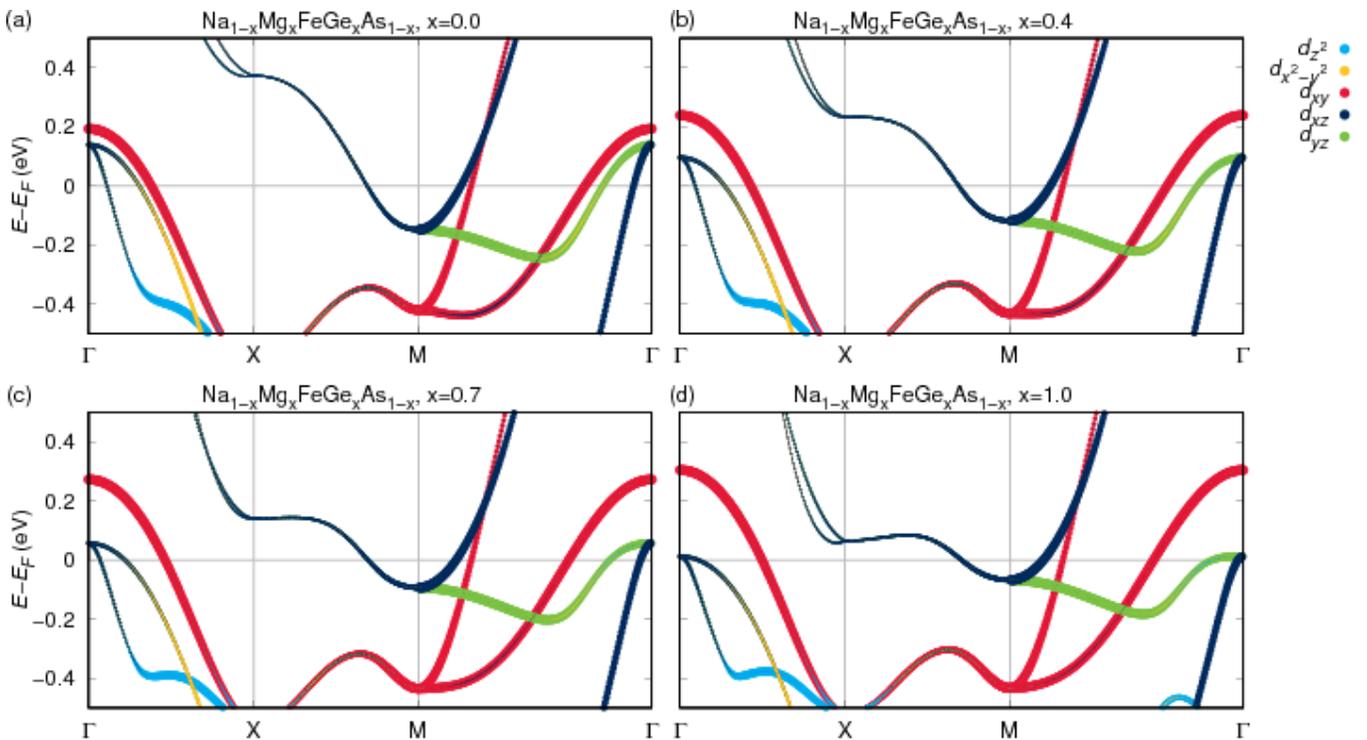

FIG. 5. Electronic bandstructure for the VCA interpolation between NaFeAs and MgFeGe. Colors indicate the weights of Fe $3d$ orbitals.

6

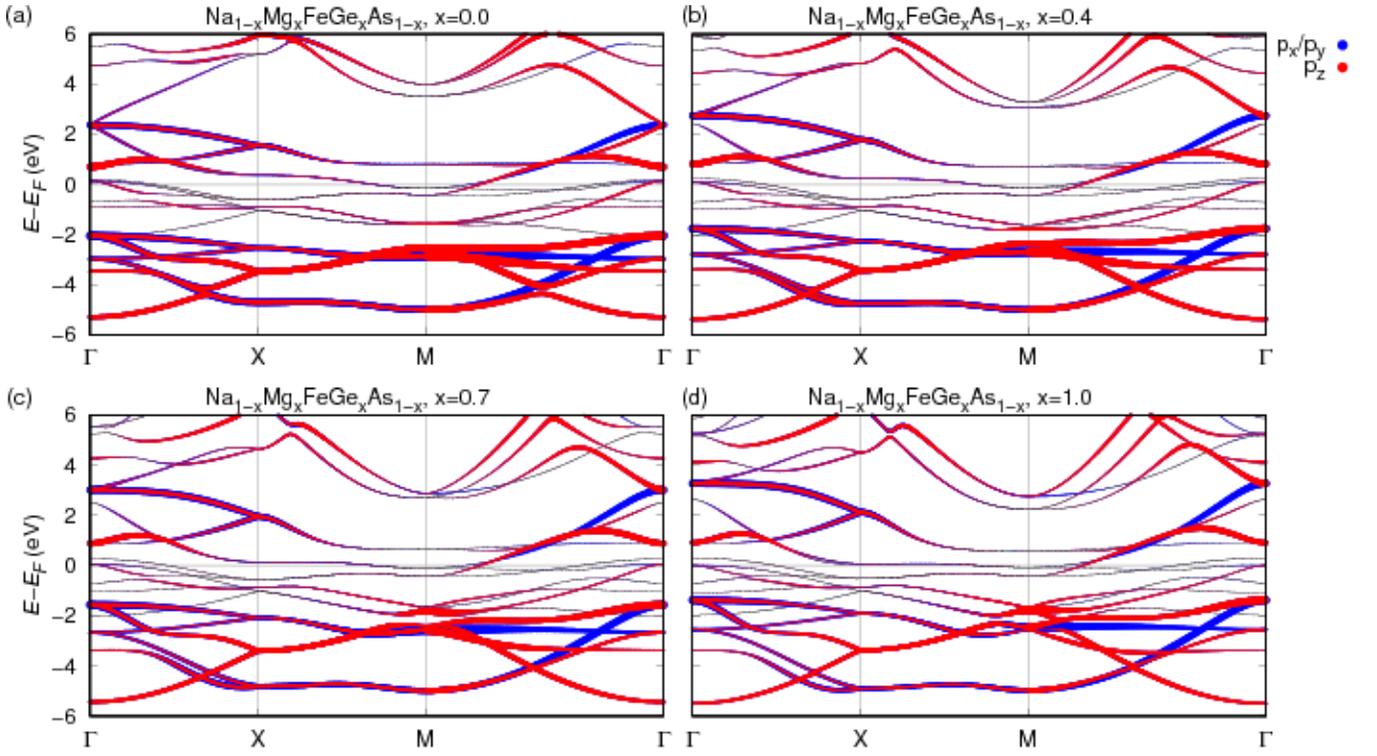

FIG. 6. Electronic bandstructure for the VCA interpolation between NaFeAs and MgFeGe. Colors indicate the weights of As/Ge 4$p$ orbitals.

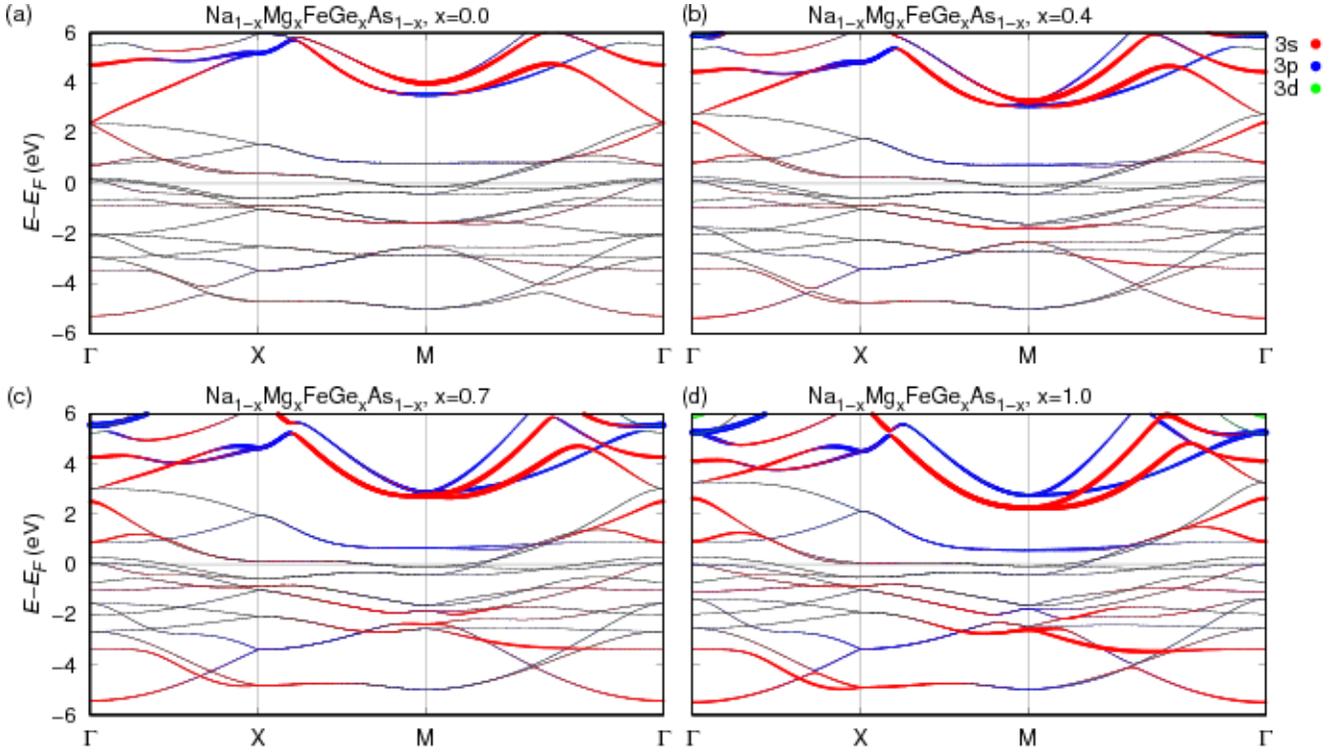

FIG. 7. Electronic bandstructure for the VCA interpolation between NaFeAs and MgFeGe. Colors indicate the weights of Na/Mg 3$s$, 3$p$ and 3$d$ orbitals.

## VI. ELECTRONIC STRUCTURE FOR VCA INTERPOLATION BETWEEN RbFe$_2$As$_2$ AND YFe$_2$Ge$_2$

We investigate the changes in electronic structure when continously turning YFe$_2$Ge$_2$ into RbFe$_2$As$_2$ using the virtual crystal approximation, while keeping all structural parameters constant. The VCA calculation is carried out starting from strontium on the yttrium/rubidium site. The electronic bandstructures with orbital weights are shown in Figs. 8, 9 and 10.

The flattening of bands described in the main text can be seen in Fig. 8. Here it is most pronounced as a flattening of Fe $3d_{xz/yz}$ states on the $M$-$\Gamma$ path segment. The effect of cation substitution on the position of the As/Ge $4p_z$ at the $\Gamma$-point is very strong here, as can be seen in Fig. 9 at the lower end of the energy window. Fig. 10 shows that the cation states are lowered very much in energy, for example at the $\Gamma$-point. Furthermore, their weight at the Fermi level strongly increases when going from RbFe$_2$As$_2$ to YFe$_2$Ge$_2$.

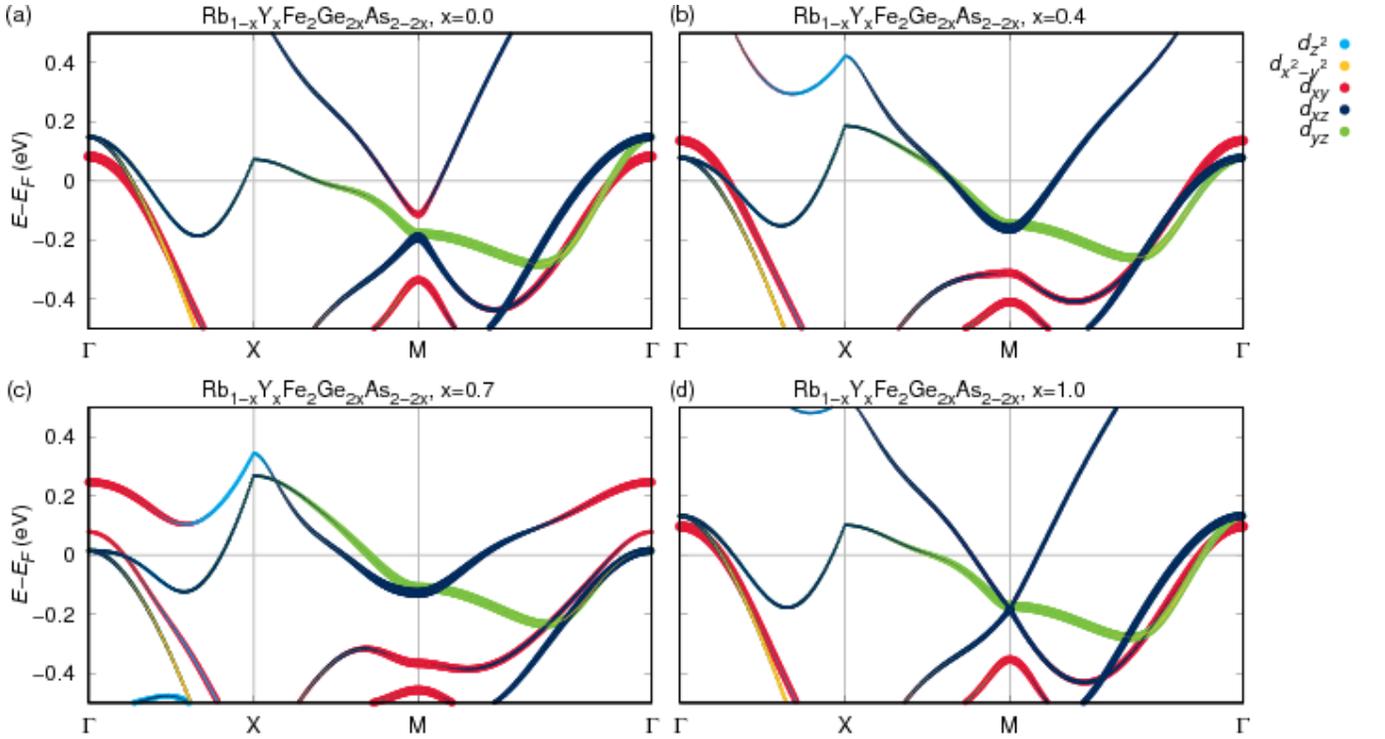

FIG. 8. Electronic bandstructure for the VCA interpolation between RbFe$_2$As$_2$ and YFe$_2$Ge$_2$. Colors indicate the weights of Fe $3d$ orbitals.

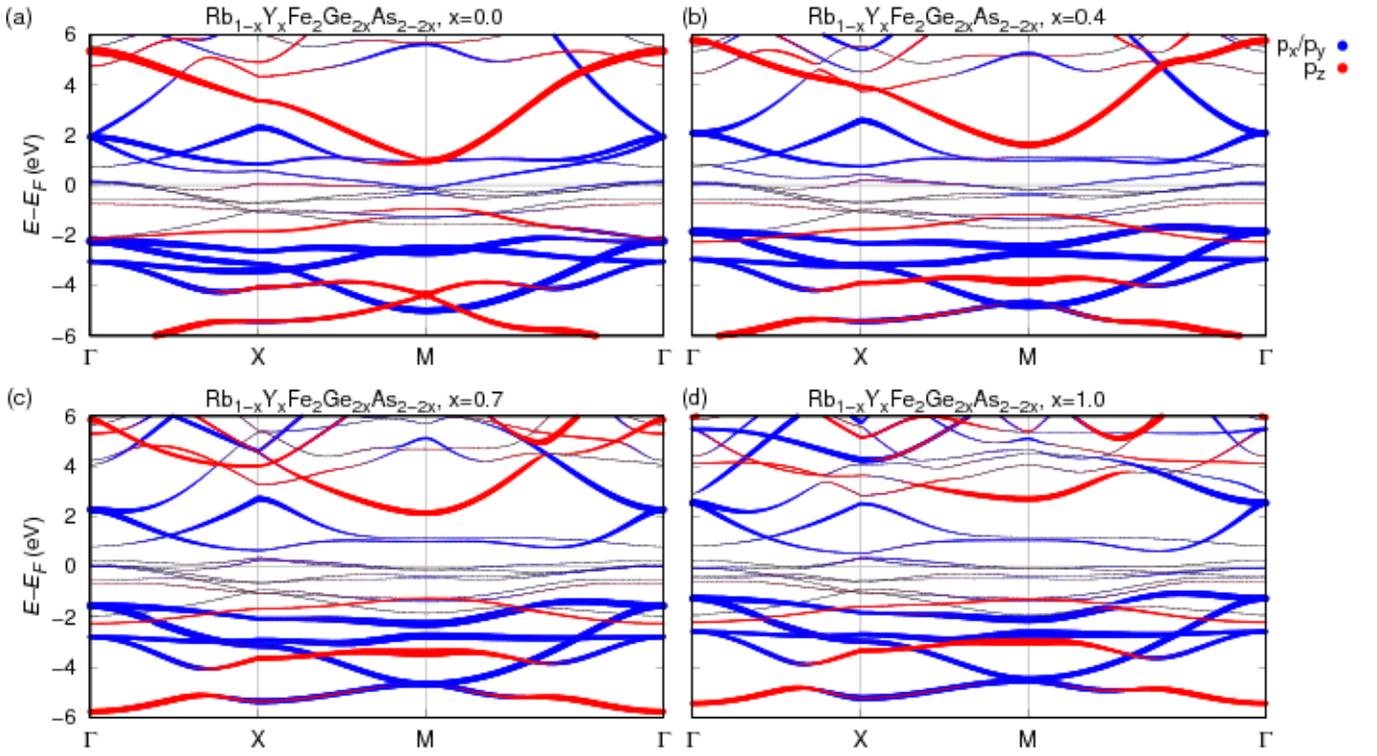

FIG. 9. Electronic bandstructure for the VCA interpolation between RbFe$_2$As$_2$ and YFe$_2$Ge$_2$. Colors indicate the weights of As/Ge $4p$ orbitals.

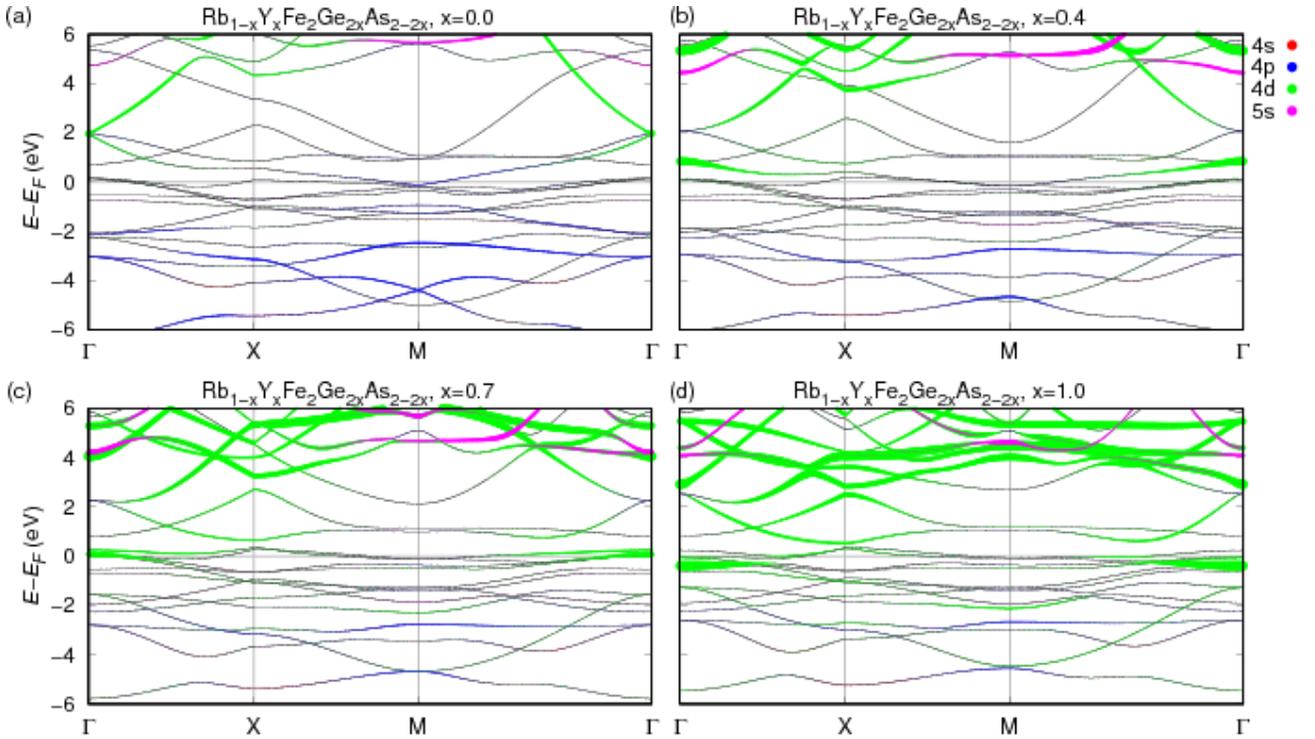

FIG. 10. Electronic bandstructure for the VCA interpolation between RbFe$_2$As$_2$ and YFe$_2$Ge$_2$. Colors indicate the weights of Rb/Y $4s$ and $4d$ orbitals.



\end{document}